\documentclass[journal=jpclcd,manuscript=letter,layout=twocolumn]{achemso}

\RequirePackage{fix-cm}
\usepackage[fontsize=11pt]{fontsize}
\usepackage[version=4]{mhchem}
\usepackage{bm}
\usepackage{amsmath,amssymb}
\usepackage{xcolor}
\definecolor{linkcolor}{HTML}{B05252}
\definecolor{citecolor}{HTML}{2563EB}
\definecolor{urlcolor}{HTML}{2563EB}
\usepackage[colorlinks,linkcolor=linkcolor,citecolor=citecolor,urlcolor=urlcolor]{hyperref}
\usepackage[capitalise]{cleveref}
\usepackage{enumitem}

\DeclareMathOperator*{\argmin}{arg\,min}

\author{Juno Nam}
\affiliation{Department of Materials Science and Engineering, Massachusetts Institute of Technology, Cambridge, Massachusetts 02139, United States}
\alsoaffiliation{Energy Storage Research Alliance, Argonne National Laboratory, 9700 South Cass Avenue, Lemont, Illinois 60439, United States}
\author{Miguel Steiner}
\affiliation{Department of Materials Science and Engineering, Massachusetts Institute of Technology, Cambridge, Massachusetts 02139, United States}
\author{Max Misterka}
\affiliation{Department of Mathematics, Massachusetts Institute of Technology, Cambridge, Massachusetts 02139, United States}
\author{Soojung Yang}
\affiliation{Computational and Systems Biology Program, Massachusetts Institute of Technology, Cambridge, Massachusetts 02139, United States}
\author{Avni Singhal}
\author{Rafael G{\'o}mez-Bombarelli}
\affiliation{Department of Materials Science and Engineering, Massachusetts Institute of Technology, Cambridge, Massachusetts 02139, United States}
\email{rafagb@mit.edu}

\title{Transferable Learning of Reaction Pathways from Geometric Priors}

\begin{document}

\begin{tocentry}
\includegraphics[width=2in,height=2in]{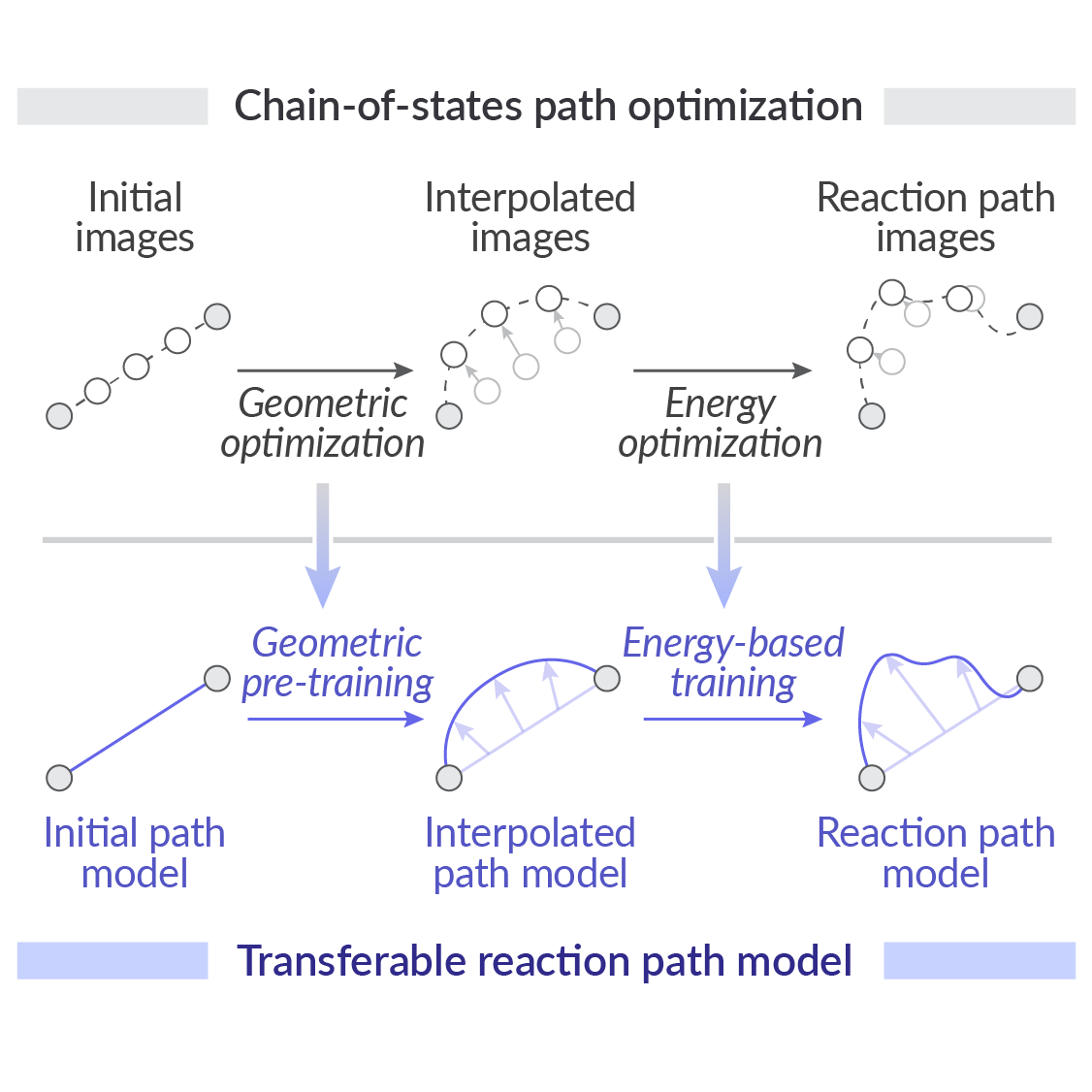}
\end{tocentry}

\begin{abstract}
Identifying minimum-energy paths (MEPs) is crucial for understanding chemical reaction mechanisms but remains computationally demanding.
We introduce MEPIN, a scalable machine-learning method for efficiently predicting MEPs from reactant and product configurations, without relying on transition-state geometries or pre-optimized reaction paths during training.
The task is defined as predicting deviations from geometric interpolations along reaction coordinates.
We address this task with a continuous reaction path model based on a symmetry-broken equivariant neural network that generates a flexible number of intermediate structures.
The model is trained using an energy-based objective, with efficiency enhanced by incorporating geometric priors from geodesic interpolation as initial interpolations or pre-training objectives.
Our approach generalizes across diverse chemical reactions and achieves accurate alignment with reference intrinsic reaction coordinates, as demonstrated on various small molecule reactions and [3+2] cycloadditions.
Our method enables the exploration of large chemical reaction spaces with efficient, data-driven predictions of reaction pathways.
\end{abstract}

\paragraph{Introduction}
Uncovering atomistic chemical reaction pathways enables the reliable prediction of properties such as reaction rates and product distributions that can accelerate various process optimization tasks.
A key step in this process is identifying the minimum energy path (MEP), which links reactants and products via the transition state (TS).
Conventional methods for locating MEPs, including the nudged elastic band (NEB) method \citep{mills1995reversible,jonsson1998nudged} and other chain-of-states techniques---which generate a series of intermediate configurations (or ``images'') between reactant and product states---require extensive energy and force evaluations, typically at the density functional theory (DFT) level.
Additionally, standard optimization procedures often suffer from issues such as kinks, corner-cutting, and sliding, \citep{henkelman2000improved,trygubenko2004doubly} as well as oscillations leading to poor convergence \citep{sheppard2008optimization}. 
As a result, these methods are computationally intensive and are unsuitable for large-scale reaction discovery across diverse chemical spaces.

Machine learning (ML) methods have been introduced to accelerate reaction path calculations and bypass the challenges of chain-of-states optimization.
A common approach is to replace expensive DFT energy and force evaluations with more efficient surrogate reactive ML interatomic potentials (MLIPs) \citep{yang2024machine,zhang2024exploring}, often referred to as the ``neural NEB'' method \citep{schreiner2022neuralneb,wander2024cattsunami,yuan2024analytical,marks2025efficient,anstine2025transferable}.
While these models significantly reduce computational costs compared to explicit quantum mechanical calculations, training reliable reactive MLIPs requires datasets of transition-region configurations, ideally sampled near the low-energy path.
This creates a ``chicken-and-egg'' problem: transition path sampling is needed to generate training data, yet the reactive MLIPs are intended to enable such sampling.
One potential solution is to use active learning strategies to improve an MLIP through iterative molecular dynamics simulations \citep{kaser2020reactive,ang2021active,li2021machine,westermayr2022deep,young2021a,young2022reaction,yang2022using,celerse2024capturing,zhang2024modelling}.
However, these strategies have only been demonstrated for individual systems, lack transferability to other reactions, and require substantial reference data.
They also rely on defining a reaction coordinate to accelerate simulations, which demands expert intuition and detailed knowledge of the reaction mechanism.
Another ML-based approach avoids conventional MEP optimizations and instead directly predicts the TS structure from the reactant and product structures. \citep{choi2023prediction,duan2023accurate,kim2024diffusion,duan2024react,zhao2025harnessing}.
At inference, this method is typically much faster than neural NEB approaches as it avoids structure optimizations.
However, the training of such a method still relies on a dataset of pre-optimized TS structures on the potential energy surface of interest, and such datasets remain limited in both size and chemical diversity.\citep{zhao2025harnessing}

\begin{figure*}[!ht]
\includegraphics[width=\textwidth]{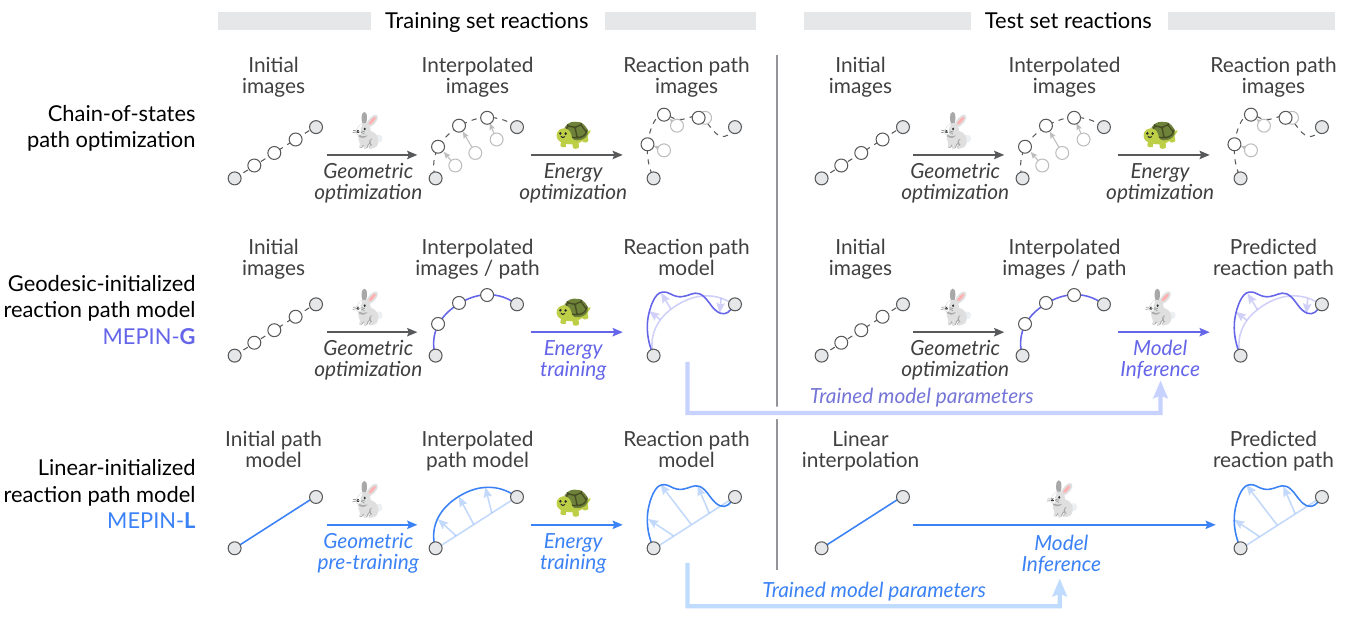}
\caption{
Schematic of the MEPIN (MEP Inference Network) training and inference.
The model leverages an implicit reaction path parametrized by a neural network to transfer knowledge of the reaction path learned during training to unseen reactant-product pairs.
It learns the difference between the minimum energy path and initial interpolation, derived from either geodesic (MEPIN-\textbf{G}) or linear (MEPIN-\textbf{L}) interpolations.
The linear-initialized model may also benefit from geometric pre-training on geodesic interpolation.
The proposed approach speeds up reaction path finding by reducing the need for costly energy evaluations at inference time.
}
\label{fig:scheme}
\end{figure*}

To address the common limitation of requiring transition pathway or structure data to model the same, we propose an alternative paradigm based on implicit parametrization of reaction pathways.
Several recent works have demonstrated that neural representations of reaction pathways, either explicitly coupled with dynamics in transition path sampling \citep{kang2024computing,singh2024splitting,du2024doobs,wang2024generalized,seong2025transition} or used as parametrized approximations of the MEP \citep{han2024stringnet,ramakrishnan2025implicit,blau2025neural}, can be effective for reaction path searches within individual systems of interest.
Applications in chemical reaction path finding have shown that continuous parametrization effectively resolves the previously mentioned issues associated with NEB optimization \citep{ramakrishnan2025implicit}.
However, these approaches are typically not designed to generalize across systems.
Here, we extend this idea to develop a chemically transferable model of a parametrized reaction path, MEPIN (MEP Inference Network), that can be trained using only reactant and product configurations along with energy and force evaluations, and then can be directly applied to unseen reactions at inference time without energy or force evaluations.
As illustrated in \cref{fig:scheme}, parametrizing the path itself eliminates the need for pre-computed transition data by generating the required information on the fly in the training process.
This removes reliance on prior transition knowledge, such as successfully optimized structures or paths, and enables a scalable and transferable framework for reaction path discovery.

\paragraph{Parametrization of reaction path}
Given the reactant configuration $x_\text{R} \in \mathbb{R}^{N \times 3}$ and product configuration $x_\text{P} \in \mathbb{R}^{N \times 3}$, each with $N$ atoms in the same index ordering, and elemental identities $a \in \mathcal{A}^N$ from a set of elements $\mathcal{A}$, we define a reaction path model $f(x_\text{R}, x_\text{P}, a, t; \theta)$ that represents the MEP between them as a function of reaction progress $t \in [0, 1]$:
\begin{align}
f(x_\text{R}, x_\text{P}, a, t; \theta) &= f_\text{interp}(x_\text{R}, x_\text{P}, a, t) \nonumber \\
& \hspace{1em} + t(1 - t)\phi(x_\text{R}, x_\text{P}, a, t; \theta),
\label{eq:rxn_path}
\end{align}
where $\phi$ is a neural network model with parameters $\theta$ that predicts the deviation from the initial interpolation $f_\text{interp}$ to the true MEP.
$f_\text{interp}$ can be any interpolation that depends only on $x_\text{R}$ and $x_\text{P}$, such as a linear interpolation $(1 - t)x_\text{R} + tx_\text{P}$ or a geodesic interpolation, as shown later.
We refer to the models as MEPIN-\textbf{L} and MEPIN-\textbf{G}, respectively (see \cref{fig:scheme}).
We explicitly parametrize the model based on the geometries and elemental identities of the reactants and products to enable chemical transferability.
In all experiments, $x_\text{R}$ and $x_\text{P}$ were translationally and rotationally aligned to minimize the root mean square deviation (RMSD).
During training, the product structures are optionally augmented by small rotations to enhance robustness (see the Supporting Information).

We implement $\phi$ as a message-passing equivariant graph neural network, adapted from the PaiNN architecture \citep{schutt2021equivariant}.
The input atomic graph is constructed with atoms as nodes and edges connecting atoms within a cutoff distance in either $x_\text{R}$ or $x_\text{P}$.
The node features encode elemental identities and the edge features are derived from interatomic distances and orientations of $x_\text{R}$, $x_\text{P}$, and the interpolated geometry $f_\text{interp}(x_\text{R}, x_\text{P}, a, t)$.

\begin{figure}[!ht]
\includegraphics[width=\columnwidth]{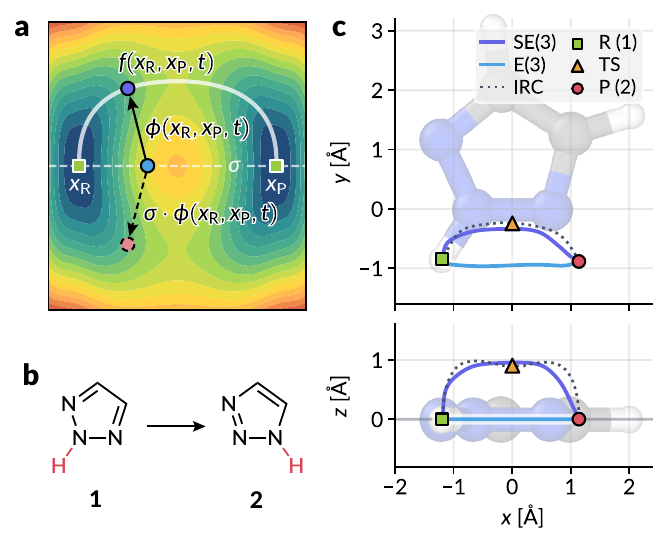}
\caption{
Symmetry considerations for the reaction path model.
(a) Two-dimensional example showing a symmetric potential landscape and linear interpolation path with respect to reflection $\sigma$, while the MEP is asymmetric. The NN model $\phi$, which learns the deviation between the initial interpolation and the MEP (scale factor $t(1-t)$ and argument $a$ are omitted), must be non-equivariant with respect to $\sigma$ to properly break the symmetry.
(b) Conversion between triazole tautomers \textbf{1} and \textbf{2}.
(c) Projections of the intrinsic reaction coordinates (IRCs) and reaction paths on the $x$--$y$ and $x$--$z$ planes learned by an SE(3)- and E(3)-equivariant neural networks for the reaction shown in panel (b). Curves and markers denote the positions of migrating hydrogen atoms, with the ring atoms aligned. The E(3) model fails to capture the out-of-plane atomic displacement.
}
\label{fig:symmetry}
\end{figure}

Message passing schemes for atomistic systems are typically constructed to preserve the physical symmetries of the potential energy function through $\mathrm{E}(3)$-equivariance.
For the reaction path in \cref{eq:rxn_path}, this implies that a group operation $g \in \mathrm{E}(3)$ transforms the path in the same way as the endpoint geometries: $f(g \circ x_\text{R}, g \circ x_\text{P}, a, t; \theta) = g \circ f(x_\text{R}, x_\text{P}, a, t; \theta)$.
However, even if the underlying potential energy is symmetric under a group, the MEP may exhibit reduced symmetries \citep{vanleeuwen2015antisymmetry}.
For example, consider a two-dimensional potential shown in \cref{fig:symmetry}a, where the potential energy, configurations $x_\text{R}$ and $x_\text{P}$, and their linear interpolants are all symmetric with respect to the mirror plane $\sigma$.
If the true path is asymmetric, as shown in \cref{fig:symmetry}a, an $\mathrm{E}(3)$-equivariant model cannot reproduce this asymmetry from symmetric inputs, as its outputs are inherently constrained to maintain planar symmetry:
\begin{align}
\sigma \circ \phi(x_\text{R}, x_\text{P}, a, t) &= \phi(\sigma \circ x_\text{R}, \sigma \circ x_\text{P}, a, t) \nonumber \\
&= \phi(x_\text{R}, x_\text{P}, a, t).
\label{eq:sym}
\end{align}
For example, in the reaction shown in \cref{fig:symmetry}b from the Transition1x set \citep{schreiner2022transition1x}, the planar molecule temporarily adopts a non-planar geometry as a hydrogen atom shifts out of the ring plane during the reaction.
An $\mathrm{E}(3)$-equivariant path model cannot capture such asymmetric displacements, resulting in a strictly planar reaction path (\cref{fig:symmetry}c).
To address this, we intentionally break the parity symmetry in our model.
Specifically, we adopt the message passing scheme from \citet{schreiner2023implicit}, which incorporates vector products to enable chiral vector predictions, making the model $\mathrm{SE}(3)$-equivariant rather than $\mathrm{E}(3)$-equivariant.
We note that while symmetry-breaking requirements may vary with the reactants, this approach can be generalized, with arbitrary symmetry reduction incorporated into the network via symmetry breaking sets \citep{xie2024equivariant}.
Additional details on the model architecture are reported in the Supporting Information.

\paragraph{Energy-based training objective}
Given the parametrized reaction path model \cref{eq:rxn_path}, we introduce an energy-based objective to optimize the parameters so that the predicted path is aligned with the unknown true MEP during training.
We leverage the variational formulation of the MEP based on the maximum reactive flux formalism \citep{berkowitz1983diffusion,huo1997maxflux}.
For a potential energy function $U: \mathbb{R}^{N \times 3} \to \mathbb{R}$, which implicitly depends on elemental identities $a$, and an arbitrary reaction path $x: [0, 1] \to \mathbb{R}^{N \times 3}$ connecting the reactant $x_\text{R} = x(0)$ and product $x_\text{P} = x(1)$ configurations, the reactive flux under overdamped Langevin dynamics is inversely proportional to the following path functional:
\begin{equation}
\mathcal{L}_\text{flux}[x(t)] = \frac{1}{\beta} \log \int_0^1 \exp \left( \beta U(x(t)) \right) \left\Vert \dot{x}(t) \right\Vert \, \mathrm{d}t,
\label{eq:flux_objective}
\end{equation}
where $\beta = 1 / k_\text{B}T$ is the inverse temperature.
Notably, in the zero-temperature limit ($\beta \to \infty$), the path minimizing \cref{eq:flux_objective} (maximizing the flux) coincides with the MEP:
\begin{equation}
x_\text{MEP}(t) = \lim_{\beta \to \infty} \argmin_{x(t)} \mathcal{L}_\text{flux}[x(t)].
\label{eq:mep}
\end{equation}
Thus, the MEP can be approximated by optimizing \cref{eq:flux_objective} at sufficiently large but numerically feasible $\beta$; in this work, we used $\beta = 20$ eV$^{-1}$.
This method, also known as the MaxFlux approach, has been successfully applied in reaction path finding using chain-of-states representation, demonstrating improved performance compared to NEB calculations \citep{koda2024flat,koda2024locating}.
Additionally, it has been employed in a proof-of-principles study on ML-based path finding on low-dimensional potentials \citep{han2024stringnet}.
We note that an alternative approach is to formulate an implicit version of NEB by stopping the gradient in directions orthogonal to the current path, as proposed in concurrent work by \citet{ramakrishnan2025implicit}.

Given a dataset of paired reactant and product geometries, we transform the variational objective in \cref{eq:flux_objective} into a loss function for the parametrized reaction path by approximating the integral with Monte Carlo sampling over structures sampled from the predicted reaction path:
\begin{equation}
\mathcal{L}_\text{flux}(\theta) = \frac{1}{\beta} \log \mathbb{E}_{t \sim \mathcal{U}(0, 1)} \left[ \exp(\beta U(x_\theta(t))) \left\Vert \dot{x}_\theta(t) \right\Vert \right],
\label{eq:flux_loss}
\end{equation}
where $x_\theta(t) = f(x_\text{R}, x_\text{P}, a, t; \theta)$ denotes an image sampled from the parametrized reaction path.

Since \cref{eq:flux_loss} does not involve derivatives of $U$, computing its gradient $\nabla_\theta \mathcal{L}_\text{flux}(\theta)$ requires only first-order derivatives $\nabla_x U$, avoiding the need for Hessian evaluations.
This makes the optimization process computationally efficient and compatible with a broad range of potential energy functions, including DFT, semiempirical quantum calculations, MLIPs, and classical force fields.

We additionally incorporate arc-length regularization \citep{han2024stringnet} to maintain a uniform distribution of points along the path.
The regularization objective is designed to keep $\Vert \dot{x}(t) \Vert$ constant along the path, and defined as:
\begin{align}
\mathcal{L}_\text{arc}[x(t)] = & \, \frac{1}{2} \int_0^1 \left( \partial_t \left\Vert \dot{x}(t) \right\Vert^2 \right)^2 \, \mathrm{d}t \nonumber \\
= & \int_0^1 \left< \dot{x}(t), \ddot{x}(t) \right>^2 \, \mathrm{d}t, \label{eq:arc_objective} \\
\mathcal{L}_\text{arc}(\theta) = & \, \mathbb{E}_{t \sim \mathcal{U}(0, 1)} \left[ \left< \dot{x}_\theta(t), \ddot{x}_\theta(t) \right>^2 \right].
\label{eq:arc_loss}
\end{align}
Since \cref{eq:flux_objective,eq:flux_loss} are invariant under reparametrization of the interpolation parameter $t$, this additional objective helps to determine an appropriate curve parametrization.
The final energy-based training loss function is a weighted sum of \cref{eq:flux_loss} and \cref{eq:arc_loss}:
\begin{equation}
\mathcal{L}_\text{energy}(\theta) = \mathcal{L}_\text{flux}(\theta) + w_\text{arc} \mathcal{L}_\text{arc}(\theta),
\label{eq:energy_loss}
\end{equation}
with $w_\text{arc}$ being a hyperparameter.

\paragraph{Geodesic initialization and pre-training}
While linear interpolation is the simplest choice for the initial interpolation, it places the model far from the MEP, leading to high computational cost during energy-based training due to a large number of potential energy evaluations.
In chain-of-states approaches, it is common to use informed initializations, such as image-dependent pair potentials (IDPP) \citep{smidstrup2014improved} or geodesic interpolations \citep{zhu2019geodesic}, which provide fast, energy-free geometric optimization serving as a good initial guess for the MEP.
We have also previously found that such interpolations provide a useful way to represent transition progress, serving as informative collective variables in the context of enhanced sampling \citep{yang2024learning}.
Similarly, we can accelerate reaction path training either by pre-training on geometric (energy-free) objective based on geodesic formalism to initialize the model closer to the MEP (MEPIN-\textbf{L}) or by directly using the geodesic path as the initial interpolation (MEPIN-\textbf{G}).
This requires a differentiable geodesic path representation with respect to $t$, which we derive based on the geodesic interpolation method by \citet{zhu2019geodesic}.

\begin{figure}[!ht]
\includegraphics[width=0.9\columnwidth]{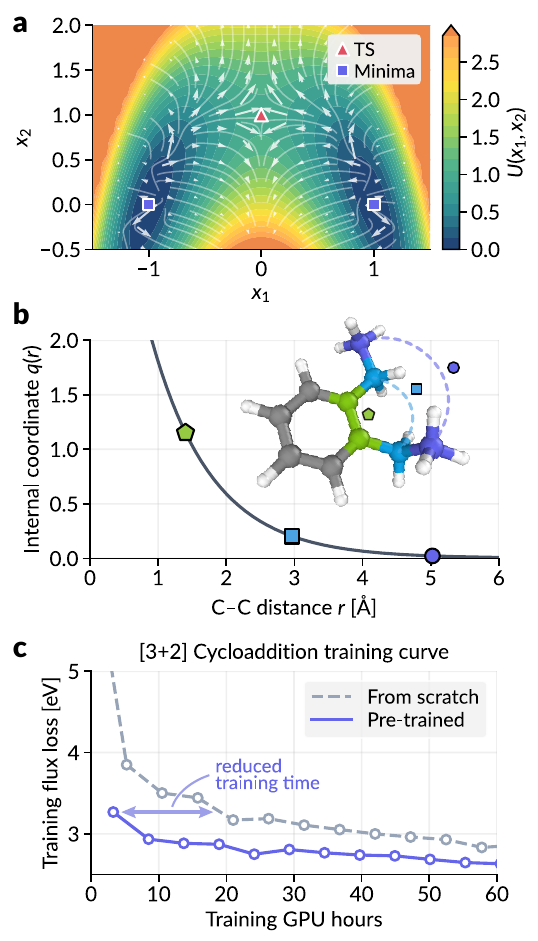}
\caption{
Geodesic interpolation.
(a) Vectors with equal ``length'' under the metric $\frac{\partial U}{\partial x^1} \frac{\partial U}{\partial x^2}$, plotted on a two-dimensional potential $U(x^1, x^2)$. The length scale contracts compared to Cartesian coordinates as the position rises on the potential surface, making the MEP the shortest path on this metric.
(b) Illustration of internal coordinates used to define the ``length'' in the geodesic interpolation scheme (\cref{eq:internal_coordinates}) for C--C interactions. Converting to internal coordinates prioritizes changes in bonded pairs over long-range non-bonded pairs.
(c) Pre-training with geodesic loss (\cref{eq:geodesic_loss}) reduces the initial flux loss (\cref{eq:flux_loss}) significantly, accelerating the overall training time (including pre-training).
}
\label{fig:geodesic}
\end{figure}

Geodesic interpolation is motivated by the differential geometry interpretation of intrinsic reaction coordinates (IRCs) by Tachibana and Fukui \citep{tachibana1978differential,tachibana1979intrinsic}.
In this framework, the metric tensor $\frac{\partial U}{\partial x^i}\frac{\partial U}{\partial x^j}$ redefines distance to effectively flatten the potential energy landscape, allowing the shortest path, i.e., the geodesic, to align with the MEP (\cref{fig:geodesic}a).
To approximate this metric without explicitly using the potential energy $U$, \citet{zhu2019geodesic} introduce a set of redundant internal coordinates $q^k$ for each atomic pair $k = (m, n)$, representing pairwise interactions, defined as:
\begin{equation}
q^{k} = \exp \left( -\bar{\alpha} \frac{r_{mn} - r_{mn}^\circ}{r_{mn}^\circ} \right) + \bar{\beta} \frac{r_{mn}^\circ}{r_{mn}},
\label{eq:internal_coordinates}
\end{equation}
where $r_{mn} = \Vert x_n - x_m \Vert$ is the interatomic distance, $r_{mn}^\circ$ is the sum of the covalent radii, and $\bar{\alpha} = 1.7$, $\bar{\beta} = 0.01$ are hyperparameters controlling short- and long-range interactions.
The infinitesimal distance element is defined through the induced metric tensor $g_{ij}$ from the coordinate transformation:
\begin{equation}
\mathrm{d}s^2 = g_{ij} \, \mathrm{d}x^i \mathrm{d}x^j = \frac{\partial q^k}{\partial x^i} \frac{\partial q^l}{\partial x^j} \delta_{kl} \, \mathrm{d}x^i \mathrm{d}x^j,
\end{equation}
where Einstein summation notation is applied to sum over repeated indices in tensor expressions, and $\delta_{kl}$ denotes the Kronecker delta.
The original geodesic interpolation is achieved by iteratively minimizing the length between the two endpoint geometries $\ell[x(t)] = \int_0^1 (g_{ij} \dot{x}^i(t) \dot{x}^j(t))^{1/2} \, \mathrm{d}t = \int_0^1 \Vert \dot{q}(t) \Vert \, \mathrm{d}t$ for a specified chain of images.
As illustrated in \cref{fig:geodesic}b, the internal coordinates (\cref{eq:internal_coordinates}) effectively amplify changes in short-range distances, causing the geodesic path to avoid atomic clashes and follow a low-energy path.

We introduce two approaches inspired by this idea: (1) using the optimized geodesic path as the initial interpolation (MEPIN-\textbf{G}) or (2) using the linear initial interpolation but pre-training the path model with a geometric objective based on geodesic interpolation before energy-based training (MEPIN-\textbf{L}).
This geometric pre-training moves the model closer to the MEP than random initialization, improving the efficiency of subsequent training.
For the former, we generate the geodesic interpolation with a predefined number of images, and then represent the continuous initial interpolation $f_\text{interp}(x_\text{R}, x_\text{P}, a, t)$ in \cref{eq:rxn_path} as a cubic B-spline \citep{bartels1995introduction} with the geodesic images serving as control points.
For the latter, we adapt the pre-training loss function to minimize the \textit{geometric} energy, $\mathcal{E}[x(t)] = \frac{1}{2} \int_0^1 \Vert \dot{q}(t) \Vert^2 \, \mathrm{d}t$, rather than the total curve length.
This improves numerical stability while preserving the path shape.
The corresponding geodesic loss for parameter optimization is then defined as:
\begin{equation}
\mathcal{L}_\text{geodesic}(\theta) = \mathbb{E}_{t \sim \mathcal{U}(0, 1)} \left[ \Vert \dot{q}_\theta(t) \Vert^2 \right].
\label{eq:geodesic_loss}
\end{equation}
We restrict the internal coordinate pairs in \cref{eq:internal_coordinates} to atom pairs within a predefined cutoff distance in the atomic graph.

In \cref{fig:geodesic}c, we demonstrate the efficiency gain of geodesic pre-training for MEPIN-\textbf{L} on the set of [3+2] cycloaddition reactions \citep{stuyver2023reaction} used in our subsequent analysis.
When plotting the flux loss (\cref{eq:flux_loss}) against the total training cost, we find that the model pre-trained with the geodesic loss starts from a much lower loss value than when trained from scratch, reducing the overall computational cost.
This shows that geodesic pre-training effectively aligns the reaction path model closer to the MEPs of the training reactions.

\begin{figure*}[!ht]
\includegraphics[width=\textwidth]{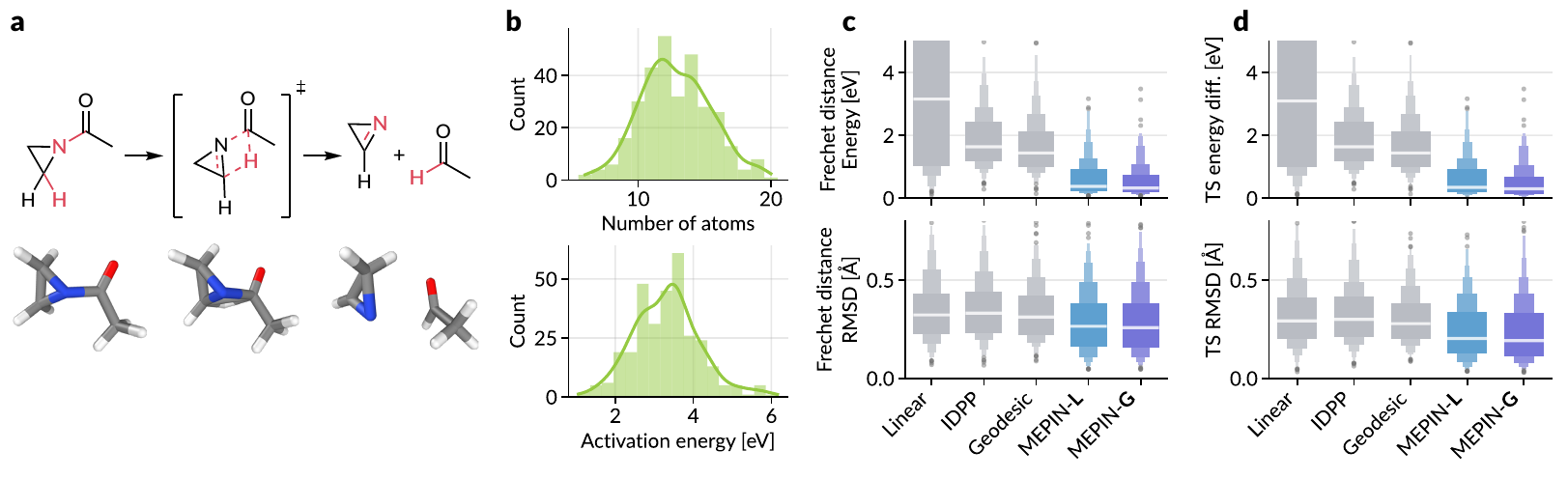}
\caption{
Results reaction path learning on the Transition1x reaction set \citep{schreiner2022transition1x}.
(a) Example reaction from the Transition1x dataset, comprising molecules with up to seven heavy atoms (C, N, O).
(b) Distribution of atom count and reaction activation energy at the GFN1-xTB level \citep{grimme2017robust} for test set reactions.
(c) Comparison of discrete Fr\'echet distance (\cref{eq:frechet_distance}) to the IRC based on potential energy difference and RMSD of different interpolation methods and reaction path models (MEPIN-\textbf{L} and \textbf{G}).
(d) Potential energy difference and RMSD between predicted and actual TS (highest energy configuration on the path), evaluated for various interpolation methods and reaction path models. The white horizontal lines represent median values.
}
\label{fig:t1x_results}
\end{figure*}

\begin{figure*}[!ht]
\includegraphics[width=\textwidth]{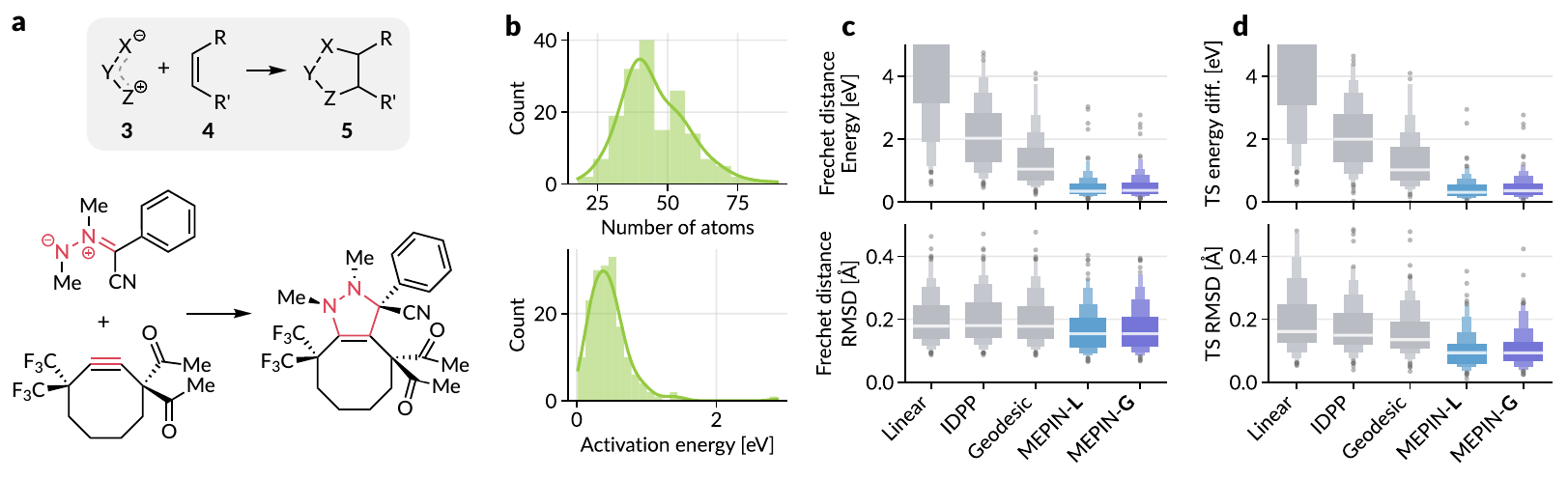}
\caption{
Results for reaction path learning on the [3+2] cycloaddition reaction set \citep{stuyver2023reaction}.
(a) General reaction scheme for [3+2] cycloadditions, where dipole \textbf{3} and dipolarophile \textbf{4} undergo 1,3-cycloaddition to form cycloadduct \textbf{5}.
An example reaction from the dataset is shown in the lower panel, with the reactive five-atom moieties highlighted.
(b) Distribution of atom count and reaction activation energy at the GFN1-xTB level \citep{grimme2017robust} for test set reactions.
(c) Comparison of discrete Fr\'echet distance (\cref{eq:frechet_distance}) to the IRC based on energy difference and RMSD of different interpolation methods and reaction path models (MEPIN-\textbf{L} and \textbf{G}).
(d) Energy difference and RMSD between predicted and actual TS (highest energy configuration on the path), evaluated for various interpolation methods and reaction path models. The white horizontal lines represent median values.
}
\label{fig:cyclo_results}
\end{figure*}

\paragraph{Reaction datasets}
To evaluate the proposed reaction path prediction approach, we created two reaction datasets derived from the Transition1x dataset \citep{schreiner2022transition1x} and the [3+2] cycloaddition reaction dataset \citep{stuyver2023reaction}.
While the method is compatible with any potential energy function that provides gradients, we employed GFN1-xTB \citep{grimme2017robust,friede2024dxtb} for computational efficiency.
The Transition1x dataset \citep{schreiner2022transition1x}, derived from \citet{grambow2020reactants}, was generated by enumerating possible elementary reactions between structures in the GDB7 dataset \citep{ruddigkeit2012enumeration}, which contains molecules with up to seven C, N, and O atoms (see example in \cref{fig:t1x_results}a).
The [3+2] cycloaddition dataset \citep{stuyver2023reaction} consists of 1,3-dipolar cycloaddition reactions, as illustrated in \cref{fig:cyclo_results}a, featuring diverse dipoles, dipolarophiles, and their substituents.
The Transition1x dataset contains small molecules with diverse reaction types and a wide range of activation barriers (\cref{fig:t1x_results}b), while the cycloaddition dataset contains larger molecules with a single reaction type, differing in reactive moieties and substituents, with a narrower range of activation barriers (\cref{fig:cyclo_results}b).

We optimized the reactant and product complexes, as well as the transition states, at the GFN1-xTB level \citep{grimme2017robust}.
To ensure compatibility with the reaction path-finding model, we retained only reactions whose reactant and product complexes preserved molecular connectivity after GFN1-xTB optimization, ensuring they remain valid on the corresponding potential energy surface.
After applying random train-test splits to these filtered datasets, we further refined the test set to include only reactions that remain unchanged on the GFN1-xTB potential energy surface with a workflow based on the \textsc{Scine} framework \citep{weymuth2024scine,vaucher2018minimum, sobez2020molassembler,molassembler2024,readuct2024,xtbwrapper2024,utilities2024}.
The verified subset of reactions includes only reactions with a transition state on the GFN1-xTB potential energy surface with only one imaginary frequency, and whose IRC connects to the optimized reactant and product complexes, as verified by \textsc{Scine Molassembler} \citep{sobez2020molassembler,molassembler2024}.
We note that for training set reactions, we verified only the endpoints, without applying TS or IRC-based filtering, assuming such information is unavailable during training.
The original TS structures in the Transition1x dataset \citep{schreiner2022transition1x} are also not IRC-verified, as they were obtained directly from climbing-image NEB (CI-NEB) calculations without further validation.
Although TS structures and NEB reaction paths were present in the original datasets, the path model was trained solely on reactant and product structures.
Further details on dataset curation and training parameters can be found in the Supporting Information.

\paragraph{Transition1x reactions}
We trained two reaction path models, MEPIN-\textbf{L} and MEPIN-\textbf{G}, on a set of 8,069 reactions in the training and validation set, and evaluated the model on 316 reactions in the test set.
For each reaction, we ran model inference with interpolation parameters $t = 0, 0.02, \cdots, 1$ to yield 51 images along the predicted path.
We compared this discretized path to the IRC connecting the same reactant and product.
We measured the Fr\'echet distance \citep{eiter1994computing} between the reference path $x(t)$ and the prediction $\hat{x}(t)$, defined as
\begin{equation}
d_F[x, \hat{x}; \delta] = \min_{\gamma, \gamma'} \max_{t \in [0, 1]} \delta \left(x(\gamma(t)), \hat{x}(\gamma'(t)) \right),
\label{eq:frechet_distance}
\end{equation}
where $\gamma, \gamma': [0, 1] \to [0, 1]$ are continuous, non-decreasing reparametrizations of $t$, and $\delta$ is a distance metric between two images.
The Fr\'echet distance measures the minimum maximum distance required to traverse both paths while maintaining respective orderings, where distances between images are measured by a distance function $\delta$.
In this work, we use two different SE(3)-invariant $\delta$: (1) the potential energy difference $\delta(x, y) = \vert U(x) - U(y) \vert$ and (2) the frame-aligned RMSD.
Additionally, since the highest-energy configuration on the MEP should correspond to the TS, we compare the highest-energy image from the discretized predicted reaction path with the reference TS structure.

The results shown in \cref{fig:t1x_results}c and \cref{fig:t1x_results}d demonstrate that both reaction path models, MEPIN-\textbf{L} and MEPIN-\textbf{G}, outperform purely geometric interpolation methods in capturing the energetic profiles of reaction pathways, with lower discrete Fr\'echet distances based on energy and smaller TS energy deviations from the reference.
The median deviations from the actual TS energies are 0.35 eV for MEPIN-\textbf{L} and 0.30 eV for MEPIN-\textbf{G}, indicating consistent and reliable identification of energetically aligned reaction paths and TSs across unseen reactions in the test set.
Although the improvements in geometry, measured by RMSD, are less pronounced, both models still perform better than the baseline interpolations.
The discrepancy between energetic and geometric alignment can be attributed to variations in molecular geometry that have minimal impact on reaction energetics or multiple reactive pathways with similar activation energies, as illustrated in the example shown in Fig. S4.
Hence, this behavior is likely due to the reaction path models being trained exclusively on energy-based objectives.

In most practical applications, the generated structures will still require optimization to confirm TS structures.
Hence, we carried out saddle-point optimizations \citep{hermes2022sella} of the predicted TSs generated by each interpolation method and reaction path model to evaluate whether the models offer improvements in both efficiency and accuracy for further refinement.
As shown in Fig. S2, the TSs predicted by the reaction path models exhibit a higher matching rate with the reference TSs after optimization and require fewer optimization steps to converge to the saddle-point structure.
This confirms that the reaction path models not only provide more accurate initial guesses but also enhance the overall computational efficiency of TS refinement workflows.

\paragraph{[3+2] cycloaddition reactions}

Similar to the Transition1x reaction set, we trained two reaction path models on 4,399 reactions from the training and validation set and evaluated them on 171 reactions in the test set.
While we used the same evaluation metrics as before, RMSD-based comparisons were computed only over the five atoms that form the five-membered ring in the cycloadduct, allowing consistent geometric evaluation across reactants with varying substituent sizes.
The results, shown in \cref{fig:cyclo_results}c and \cref{fig:cyclo_results}d, indicate that both models achieve again strong energetic alignment to the IRCs, with median TS energy deviations of 0.31 eV for MEPIN-\textbf{L} and 0.36 eV for MEPIN-\textbf{G}.

Unlike the results on Transition1x, MEPIN-\textbf{G} performs comparably to or worse than MEPIN-\textbf{L} in both energetic and geometric alignment for the [3+2] cycloaddition set.
While explicitly optimized geodesic paths may be more beneficial for datasets with diverse reaction types, in cases like the [3+2] cycloaddition set---where reactions follow a consistent mechanistic motif---the pre-training and fine-tuning strategy of MEPIN-\textbf{L} appears sufficient to replace such initialization.
In fact, the added curvature from geodesic-based paths may overly constrain the model, limiting its flexibility to adapt to patterns that can be learned directly from the data.

\begin{figure}[!ht]
\includegraphics[width=\columnwidth]{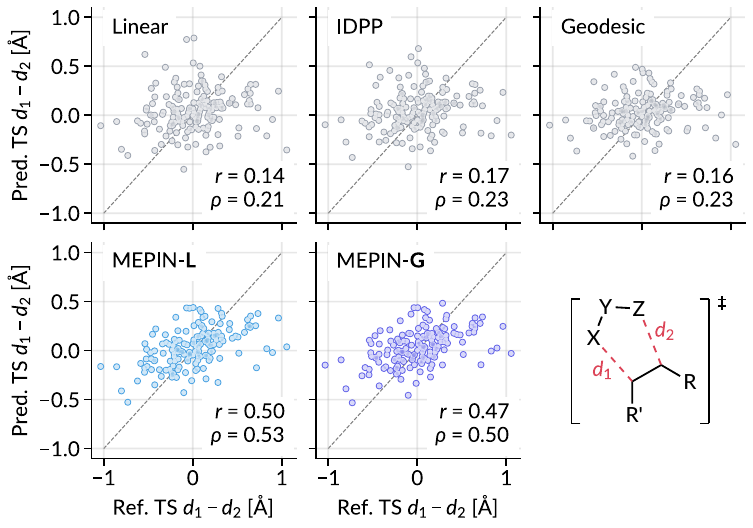}
\caption{
Probing the asynchronicity of the [3+2] cycloaddition with the reaction path model.
For each interpolation method and reaction path model, the difference between the two forming bond distances at the TS (highest energy configuration) is compared to the reference TS for the test set reactions.
Models trained with potential energy yield noticeably higher Pearson ($r$) and Spearman ($\rho$) correlations compared to purely geometric interpolation methods.
}
\label{fig:cyclo_ts_bond}
\end{figure}

For the predicted TS structures, we again carried out saddle-point optimizations \citep{hermes2022sella} to assess whether the model predictions improve computational efficiency in refinement workflows (Fig. S3).
We similarly observed that learned reaction path models reduce the cost of TS refinement, with a notable increase in cases where the optimized TS matches the reference structure.
Additionally, we analyzed the asynchronicity of the bond formation during the cycloaddition process to assess whether the predicted reaction paths capture chemical features beyond geometric interpolation.
Cycloaddition reactions proceed in a concerted yet asynchronous fashion, with the degree of asynchronicity influenced by the substituents and reactive moieties \citep{barrales2021dipolar,kawamura2022mechanistic}.
To quantify this, we compared the differences in bond lengths for the two forming bonds at the predicted versus the reference TS structures.
As illustrated in \cref{fig:cyclo_ts_bond}, the model predictions show much stronger correlations with reference asynchronicity than geometric interpolations, indicating that the models learned to capture subtle, reaction-specific nuances and generalize across substituent variations.

\paragraph{Computational cost and comparison with previous methods}
The computational costs of the reaction path models (training and inference) and interpolation methods are provided in Supporting Information Section S3.3.
Model inference takes less than 1 ms on a GPU and 10--30 ms on a CPU per predicted image, comparable to or faster than interpolation methods based on geometric optimization (e.g., IDPP and geodesic).
Although training involves a few hundred energy and force evaluations per reaction, it does not require pre-optimized TSs or reaction paths, and inference requires no energy evaluations.

While there were no prior works on transferable prediction of entire reaction paths, recent studies on ML-based transition characterization have focused on directly predicting TS structures \citep{choi2023prediction,duan2023accurate,kim2024diffusion,duan2024react,zhao2025harnessing} or pre-computed reaction energetics \citep{vangerwen2022physics,vangerwen20243dreact}.
These approaches depend on large datasets of TS structures generated via computationally intensive methods like CI-NEB, which are challenging to automate and prone to failure.
Additionally, they typically yield only a single TS structure or activation energy per inference, limiting their application to broad, mechanism-free reaction discovery.
In contrast, our method eliminates the need for explicit TS annotations by training exclusively on endpoint structures with energy-based supervision, enabling the generation of full reaction paths that can be discretized as needed.
This makes our approach well-suited for integration into automated reaction enumeration and exploration pipelines \cite{Dewyer2018,Simm2019,Unsleber2020,Baiardi2022,Ismail2022,wen2023chemical,Margraf2023,Steiner2022}.

Some of the aforementioned works reported strong performance on specific datasets---for example, \citet{duan2024react} report $\sim 0.075$ \AA{} RMSD for TS structure reproduction in the Transition1x dataset \citep{schreiner2022transition1x}.
However, such metrics may not reflect the success in recovering the correct path information, as the TS structures in the dataset are derived directly from CI-NEB calculations and are not verified by IRC calculations.
Thus, the modeling approach should be selected based on the desired prediction outcome and the availability of prior transition path or structure data.

\paragraph{Conclusion}
We have introduced MEPIN, an efficient and scalable reaction path modeling approach that accelerates the identification of MEPs without requiring prior knowledge of TS structures during training.
By combining a parametrized reaction path model with an energy-based training objective and a geodesic initialization or pre-training strategy, our approach achieves accurate energetic alignment with reference IRCs across diverse reaction types.
Evaluations on both the Transition1x and [3+2] cycloaddition reaction sets demonstrate generalization to unseen reaction pathways and features.
Future work could improve this approach by exploring transfer learning or multi-fidelity strategies across multiple potential energy surfaces to leverage high-accuracy DFT-level data \citep{greenman2022multi}.
In addition, incorporating generative modeling to discover multiple possible reaction pathways \citep{hayashi2024generative} would be crucial, as reactions can proceed via alternative routes that are not captured by current single-path modeling strategies.
Furthermore, multi-objective learning methods \citep{bonati2023unified} that additionally integrate available TS or optimized reaction path data could further improve geometric accuracy and performance, addressing the current limitations of our model compared to methods explicitly trained on known TS structures or reaction paths.

\paragraph{Data availability}
The code and data to reproduce the results of this work have been made available on GitHub: \url{https://github.com/learningmatter-mit/mepin}.

\begin{acknowledgement}
The authors thank Xiaochen Du for carefully reviewing the manuscript and providing valuable suggestions, and Minho Kim, MinGyu Choi, and Luigi Bonati for insightful discussions.
This research used resources of the National Energy Research Scientific Computing Center (NERSC), a Department of Energy Office of Science User Facility using NERSC awards BES-ERCAP-m4737 and BES-ERCAP-m4866.
We also acknowledge the MIT SuperCloud and Lincoln Laboratory Supercomputing Center for providing HPC resources.
J.N. was supported by the Energy Storage Research Alliance ``ESRA'' (DE-AC02-06CH11357), an Energy Innovation Hub funded by the U.S. Department of Energy, Office of Science, Basic Energy Sciences.
M.S. gratefully acknowledges the Mobility fellowship P500PN\_225736 from the Swiss National Science Foundation.
M.M. was supported by the MIT Undergraduate Research Opportunities Program.
A.S. acknowledges support from the National Defense Science and Engineering Graduate Fellowship.
\end{acknowledgement}

\begin{suppinfo}
Details on model architecture, training procedures, and hyperparameters;
dataset preparation and comparison with the original dataset;
and additional results, including saddle-point optimization, inference examples, and computational cost.
\end{suppinfo}

\bibliography{main}

\end{document}